\documentclass[12pt]{article}
\usepackage{graphics} 
\def\lsim{\mathrel{\lower2.5pt\vbox{\lineskip=0pt\baselineskip=0pt
          \hbox{$<$}\hbox{$\sim$}}}}
\def\gsim{\mathrel{\lower2.5pt\vbox{\lineskip=0pt\baselineskip=0pt
          \hbox{$>$}\hbox{$\sim$}}}}
\begin{document}   
%------------------------------------------------------------------------------
\markright{Brane surgery: energy conditions, traversable wormholes, and voids.}
%------------------------------------------------------------------------------
\def\Barcelo{Barcel\'o}
%------------------------------------------------------------------------------
\title{{\bf Brane surgery: energy conditions, \\ 
\bf traversable wormholes, and voids}}
\author{Carlos \Barcelo\ and Matt Visser\\[2mm]
{\small \it Physics Department, Washington University, 
Saint Louis, Missouri 63130-4899, USA.}}
\date{{\small 4 April 2000; \\
Revised 18 April 2000 (reference and footnote added); \\
Revised 8 May 2000 (terminology standardized, references added);\\
\LaTeX-ed \today}}
%------------------------------------------------------------------------------
\maketitle
%------------------------------------------------------------------------------
\begin{abstract}
Branes are ubiquitous elements of any low-energy limit of string
theory. We point out that negative tension branes violate all the
standard energy conditions of the higher-dimensional spacetime they
are embedded in; this opens the door to very peculiar solutions of the
higher-dimensional Einstein equations. Building upon the
(3+1)-dimensional implementation of fundamental string theory, we
illustrate the possibilities by considering a toy model consisting of
a (2+1)-dimensional brane propagating through our observable
(3+1)-dimensional universe. Developing a notion of ``brane surgery'',
based on the Israel--Lanczos--Sen ``thin shell'' formalism of general
relativity, we analyze the dynamics and find traversable wormholes,
closed baby universes, voids (holes in the spacetime manifold), and an
evasion (not a violation) of both the singularity theorems and the
positive mass theorem. These features appear generic to any brane
model that permits negative tension branes: This includes the
Randall--Sundrum models and their variants.

\vspace*{5mm}
\noindent
PACS: 04.60.Ds, 04.62.+v, 98.80 Hw
\\
Keywords: branes, brane surgery, energy conditions, wormholes,
voids.
\end{abstract}
%-----------------------------------------------------------------------
\vfill
%----------------------------------------------------------------------
\hrule
%-----------------------------------------------------------------------
\bigskip
%-----------------------------------------------------------------------
\centerline{\underline{E-mail:} {\sf carlos@hbar.wustl.edu}}
%-----------------------------------------------------------------------
\centerline{\underline{E-mail:} {\sf visser@kiwi.wustl.edu}}
%-----------------------------------------------------------------------
\bigskip
%-----------------------------------------------------------------------
\centerline{\underline{Homepage:} {\sf http://www.physics.wustl.edu/\~{}carlos}}
%-----------------------------------------------------------------------
\centerline{\underline{Homepage:} {\sf http://www.physics.wustl.edu/\~{}visser}}
%-----------------------------------------------------------------------
\bigskip
%-----------------------------------------------------------------------
\centerline{\underline{Archive:}
{\sf hep-th/0004022}}
%----------------------------------------------------------------------
\bigskip
\hrule
\clearpage
%---------------------------------------------------------------------
% User definitions
%--------------------------------------------------------------------
\def\Box{\nabla^2}
%---------------------------
\def\d{{\mathrm d}}
%----------------------------
\def\ie{{\em i.e.\/}}
\def\eg{{\em e.g.\/}}
\def\etc{{\em etc.\/}}
\def\etal{{\em et al.\/}}
%----------------------------
\def\S{{\mathcal S}}
\def\I{{\mathcal I}}
\def\L{{\mathcal L}}
\def\R{{\mathcal R}}
\def\eff{{\mathrm{eff}}}
\def\Newton{{\mathrm{Newton}}}
\def\bulk{{\mathrm{bulk}}}
\def\matter{{\mathrm{matter}}}
\def\tr{{\mathrm{tr}}}
\def\normal{{\mathrm{normal}}}
\def\implies{\Rightarrow}
\def\half{{1\over2}}
\def\induced{{\mathrm{induced}}}
\def\effective{{\mathrm{effective}}}
\def\Nordstrom{{Nordstr\"om}}
\def\RNdS{Reissner--\Nordstrom--de~Sitter}
%------------------------------------
\def\t{{\hat{t}}}
\def\th{{\hat{\theta}}}
%----------------------------------------------------------------------- 
\def\SIZE{1.00} % for graphics package
%-----------------------------------------------------------------------

%-----------------------------------------------------------------------
\clearpage
%-----------------------------------------------------------------------
\section{Introduction}
%------------------------------------------------------------------------------
\label{S:intro}
%------------------------------------------------------------------------------

Branes, ubiquitous elements of any low-energy limit of string
theory, have recently attracted much attention as essential
ingredients of the semi-phenomenological Randall--Sundrum
models~\cite{RS1,RS2}. These models have been used to both ameliorate
the ``hierarchy problem''~\cite{RS1} and to explore the possibility of
``exotic'' Kaluza--Klein theories with their infinitely large extra
dimensions~\cite{RS2}. Essential ingredients in these RS models are
the existence of both {\em positive} and {\em negative} tension
branes.

Now a brane tension is normally thought of as being completely
equivalent to an internal cosmological constant, and from the point of
view of physics constrained to the brane this is certainly
correct. However, from the higher-dimensional point of view (that is,
as seen from the embedding space) this is not correct: For a
(p+1)-brane embedded in (n+1) dimensions a brane tension leads
to the stress energy
\begin{equation}
T^{\mu\nu} = 
-\Lambda_D \; g^{\mu\nu}_\induced \; \delta^{n-p}(\eta^a)= 
-\Lambda_D \; 
\left(g^{\mu\nu} - \sum_{a=1}^{n-p} n_a^\mu \; n_a^\nu \right) \; 
\delta^{n-p}(\eta^a),
\end{equation}
where the sum runs over the $n-p$ normals to the brane, and the
$\eta^a$ are suitable Gaussian normal coordinates. Contracting with a
higher-dimensional null vector, $k_\mu$, we see
\begin{equation}
T^{\mu\nu} \; k_\mu \; k_\nu = 
-\Lambda_D \; g^{\mu\nu}_\induced  \; k_\mu \; k_\nu  \; 
\delta^{n-p}(\eta^a)= 
\Lambda_D \; 
\left[\sum_{a=1}^{n-p} (n_a^\mu \; k_\mu )^2 \right]\; 
\delta^{n-p}(\eta^a).
\end{equation}
If the brane tension is negative, $\Lambda_D < 0$, and the null vector
is even slightly orthogonal to the brane, then on the brane
\begin{equation}
T^{\mu\nu} \; k_\mu \; k_\nu < 0.
\end{equation}
That is, the embedding-space null energy condition (NEC) is
violated. In fact, integrating across the brane, even the averaged
null energy condition (ANEC) is violated. ({\em Ipso facto}, all the
energy conditions are violated.) This is a classical violation of the
energy conditions, which we shall soon see is even more profound than
the classical violations due to non-minimally coupled scalar
fields~\cite{Flanagan-Wald}.

In a recent series of papers~\cite{Barcelo1,Barcelo2,Barcelo3} we have
made a critical assessment of the current status of the energy
conditions, finding a variety of both classical and quantum violations
of the energy conditions. We now see that uncontrolled violations of
the energy conditions are also a fundamental and intrinsic part of any
brane-based low-energy approximation to fundamental string
theory. Among the possible consequences of these energy condition
violations we mention the occurrence of traversable wormholes
(violations of topological censorship), possible violations of the
singularity theorems (more properly, evasions of the singularity
theorems), and even the possibility of negative asymptotic mass.

A particular example of this sort of phenomenon occurs in the (finite
size) Randall--Sundrum models, where one has two parallel branes (our
universe plus a hidden brane) of equal but opposite brane tension. One
or the other of these branes (depending on whether one is considering
the RS1 or RS2 model) violates the (4+1)-dimensional energy conditions
and exhibits the ``flare out'' behaviour reminiscent of a traversable
wormhole~\cite{Anchordoqui}. That these branes do not quite represent
traversable wormholes in the usual sense~\cite{Morris-Thorne,Book}
follows from the fact that the ``throat'' is an entire flat (3+1)
Minkowski space, instead of the more usual $R^1\times
S^{d-1}$. Furthermore, in the infinite-size version of the
Randall--Sundrum (RS2) model, where the hidden sector has been pushed
out to hyperspatial infinity, our universe is itself represented by a
positive-tension (3+1)-brane, which does not violate any
(4+1)-dimensional energy conditions. The energy-condition violating
brane has in this particular model been pushed out to hyperspatial
infinity and discarded. Be that as it may, the occurrence of negative
tension branes in modern semi-phenomenological models is generic, and
a feel for the some of the peculiar geometries they can engender is
essential to developing any deep understanding of the physics.

In this particular paper we shall for illustrative purposes choose a
particularly simple model: We work with a (3+1)-dimensional bulk,
which contains a (2+1)-dimensional brane (of either positive or
negative brane tension). We choose this particular model because it is
sufficiently close to reality to make the points we wish to make as
forcefully as possible, and because it arises naturally in certain
types of fundamental string theory.  While it is most often the case
that fundamental string theories (or their various offspring: membrane
models, M-theory, \etc) are formulated in either (9+1) or (10+1)
dimensions,\footnote{
%-----------------------
In many specific cases the actual implementation is directly in terms
of a Euclidean-signature $10$ or $11$ dimensional spacetime; with the
underlying Lorentzian-signature reality hidden under several layers of
scaffolding.  
}
%-----------------------
this is not absolutely necessary: There is an entire industry based on
formulating string theories directly in (3+1) dimensions, with the
price that has to be paid being the inclusion of extra
(1+1)-dimensional quantum fields propagating on the
world-sheet~\cite{Tye1,Tye2,Tye3,Tye4,Tye5}.\footnote{
%-----------------------
Consider for example the bosonic string, which is most often viewed as
a (1+1)-dimensional world sheet propagating in (25+1) dimensions:
There is a trivial re-interpretation in which the bosonic string
propagates in (3+1) dimensions and there are $22$ free scalar fields
propagating on the world-sheet. These $22$ scalar fields are there
just to soak up the conformal anomaly and make the theory
manageable. If these scalar fields are now constrained by appropriate
identifications the re-interpretation is less trivial --- it is an
example of the fact that compactifications of {\em some} of the
dimensions of the higher-dimensional embedding spacetime that the
world sheet propagates through can be traded off for a
lower-dimensional uncompactified embedding spacetime plus interacting
fields on the world-sheet. When this procedure is applied to
superstrings the technical details are considerably more complex, but
the basic result still holds.}
%-----------------------
Now even in such a (3+1)-dimensional incarnation of string theory,
open strings will terminate on D-branes (Dirichlet branes), and an
effective theory involving the (3+1)-dimensional bulk plus (2+1),
(1+1), and (0+1) dimensional D-branes (``domain walls'', ``cosmic
strings'', and ``soliton-like particles'') can be contemplated as a
low-energy approximation.\footnote{
%------------------------
More traditional string theorists who absolutely insist on working
directly in the higher-dimensional embedding space can view the
current calculations as a particular toy model in which only selected
sub-sectors of the grand total degrees of freedom are excited.
Additionally, it should be borne in mind that many of the generic
features of the analysis presented in this paper will extend {\em
mutatis mutandis\/} to embedding spaces and branes of higher
dimensionality. You do not want the bulk to have fewer than (3+1)
dimensions since then bulk gravity is either completely or almost
trivial. You do not want the bulk to have more than (10+1)
dimensions since the model is then difficult to interpret in terms of
fundamental string theory. For technical reasons (to be able to use
the thin-shell formalism) you want the brane to be of co-dimension
$1$, so if the bulk is (n+1)-dimensional the brane should be
([n--1]+1)-dimensional. Within these dimensional limitations, the
qualitative features of this paper are generic.
}
%------------------------
While D-branes are perhaps the most straightforward examples of
membrane-like solitons in string theory, they do come with additional
technical baggage: the most elementary implementation of D-branes
occurs in bosonic string theories~\cite{Polchinski-1}, but often
D-branes are associated with specific implementations of
supersymmetric string theories~\cite{Polchinski-2} and carry various
types of Ramond--Ramond or Neveu--Schwarz charge. There are in
addition other types of brane-like configurations that sometimes arise
in fundamental string theory such as non-dynamical ``orientifold
planes''~\cite{Polchinski-2}, which generate gravitational fields
corresponding to negative tensions, but which do not themselves exhibit
internal dynamics. We will not delve further into this beastiary, but
will instead content ourselves with the observation that the
low-energy limit of fundamental string theory (of whatever persuasion)
generically leads to an effective theory containing brane-like
excitations.

This overall picture is actually very similar to the notion of extended
topological defects arising from symmetry breaking in point-particle
field theories: There are many semi-phenomenological GUT-based point
particle field theories that naturally contain domain walls, cosmic
strings, and/or solitons. The key difference here is that point
particle field theories inevitably lead to positive brane tensions,
with negative brane tensions being energetically disfavoured (they
correspond to an unnatural form of symmetry breaking that forces one
to the {\em top} of the potential). The key difference in brane-based
models is that there is no longer any particular barrier to negative
brane tension --- in fact negative brane tensions are ubiquitous, now
being so commonly used as to almost not require explicit
mention~\cite{Negative}.

Within the model we have chosen, we demonstrate that negative tension
branes lead to traversable wormholes --- in some cases to stable
traversable wormholes. (Positive tension branes quite naturally lead
to closed baby universes; these are {\em not\/} FLRW universes, and
are not suitable for cosmology, but are perhaps of interest in their
own right.) We also explore the possibility of viewing the brane as
an actual physical boundary of spacetime, with the region on the
``other side'' of the brane being null and void.

The basic tools used are the idea of ``Schwarzschild surgery'' as
developed in~\cite{Surgery} (see also the more detailed presentation
in~\cite{Book}), which we first extend to ``brane surgery'',
specialize to ``\RNdS'' surgery, and then use to present an analysis
of both static and dynamic spherically-symmetric (2+1)-dimensional
branes in a (3+1)-dimensional {\RNdS} background geometry.\footnote{
%-----------------------------------------------------------
As we shall soon see, brane surgery is essentially a specific
implementation of the Israel--Lanczos--Sen junction conditions of
general relativity; as such it has been used implicitly in many
brane-related papers (see for
example~\cite{RS1,RS2,Arkani,Chacko,Grojean}); the key difference in
the present paper is in the details and in the questions we address.}
%-----------------------------------------------------------
We find both stable and unstable traversable wormhole solutions,
stable and unstable baby universes, and stable and unstable voids.

%-----------------------------------------------------------------------------
\section{Brane surgery}
%------------------------------------------------------------------------------
\label{S:surgery}
%------------------------------------------------------------------------------

We start by considering a rather general static spherically symmetric
geometry (not the most general, but quite sufficient for our purposes)
\begin{equation}
\d s^2 = - F(r) \; \d t^2 +{\d r^2\over F(r)} + r^2 \; \d\Omega_2^2.
\end{equation}
To build the class of geometries we are interested in, we start by
taking two copies of this geometry, truncating them at some
time-dependent radius $a(t)$, and sewing the resulting geometries
together along the boundary $a(t)$. The result is a manifold without
boundary that has a ``kink'' in the geometry at $a(t)$. If we sew
together the two external regions $r\in(a(t),\infty)$, then the result
is a wormhole spacetime with two asymptotic regions. On the other
hand, if we sew together the two internal regions $r\in(0,a(t))$, then
the result is a closed baby universe.

At the ``kink'' $a(t)$ the spacetime geometry is continuous, but the
radial derivative (and hence the affine connexion) has a step-function
discontinuity. The Riemann tensor in this situation has a
delta-function contribution at $a(t)$, and this geometry can be
analyzed using the Israel--Lanczos--Sen ``thin shell'' formalism of
general relativity~\cite{Israel,Lanczos,Sen}. The relevant specific
implementation of the thin-shell formalism can be developed by
extending the formalism of~\cite{Surgery} and~\cite{Book}. Because of
its relative simplicity we shall start with the static case
$a=\mathrm{constant}$.

%----------------------------------------------------
\subsection{Brane statics}
%----------------------------------------------------
\label{SS:statics}
%---------------------------------------------------

The unit normal vector to the sphere $a=\mathrm{constant}$ is
(depending on whether one is considering inward or outward normals)
\begin{equation}
n^\mu = \pm\left(0,\sqrt{F(a)},0,0\right); \qquad 
n_\mu = \pm\left(0,{1\over\sqrt{F(a)}},0,0\right).
\end{equation}
The extrinsic curvature (second fundamental form) can be written
in terms of the normal derivative
\begin{equation}
K_{\mu\nu} 
= \half \;{\partial g_{\mu\nu}\over\partial\eta}
= \half \; n^\sigma \;  {\partial g_{\mu\nu}\over\partial x^\sigma}
= \pm \half \; \sqrt{F(a)} \;\; {\partial g_{\mu\nu}\over\partial r}.
\end{equation}
If we go to an orthonormal basis, the relevant components are\footnote{
%----------------------------------------------------------------------
The use of an orthonormal basis makes it particularly easy to phase
the calculation in terms of the physical density and physical
pressure.}
%----------------------------------------------------------------------
%
\begin{equation}
K_{\t\t} 
= \mp \half  \sqrt{F(r)} \; {\partial g_{tt}\over\partial r} \; g^{tt} 
= \mp \half  \sqrt{F(r)} \; {\partial F(r)\over\partial r} \; {1\over F(r)} 
= \mp \half  {F(r)}^{-1/2} \; \left.{\partial F(r)\over\partial r}\right|_{r=a}.
\end{equation}
\begin{equation}
K_{\th\th} 
= \pm \half  \sqrt{F(r)} \; {\partial g_{\theta\theta}\over\partial r} \; 
   g^{\theta\theta} 
= \pm \half  \sqrt{F(r)} \; {\partial r^2\over\partial r} \; {1\over r^2} 
= \pm \left.{\sqrt{F(r)}\over r}\right|_{r=a}.
\end{equation}
The discontinuity in the extrinsic curvature is related to the jump in
the normal derivative of the metric as one crosses the brane
\begin{equation}
\kappa_{\mu\nu} = K^+_{\mu\nu} - K^-_{\mu\nu}.
\end{equation}
In general, one could take the geometry on the two sides of the brane
to be different $[F^+(r)\neq F^-(r)]$, but in the interests of clarity
the present models will all be taken to have a $Z_2$ symmetry under
interchange of the two bulk regions.\footnote{
%------------------------------------------------------------------
Remember that we have already decided to take the range of the $r$
coordinate to be {\em either} two copies of $(a(t),\infty)$,
corresponding to a wormhole; {\em or} two copies of $(0,a(t))$,
corresponding to a baby universe. Then $Z_2$ symmetry corresponds to
$F^+(r)= F^-(r)$, with a kink in the geometry at $r=a(t)$. Our normal
vectors do not flip sign as we cross the brane.}
%-----------------------------------------------------------------
Under these conditions
\begin{equation}
\kappa_{\t\t} 
= \mp  {F(r)}^{-1/2} \; \left.{F(r)\over\partial r}\right|_{r=a}.
\end{equation}
\begin{equation}
\kappa_{\th\th} 
= \pm 2 \left.{\sqrt{F(r)}\over r}\right|_{r=a}.
\end{equation}
The upper sign refers to a wormhole geometry where the two exterior
regions have been sewn together (discarding the two interior regions),
while the lower sign is relevant if one has kept the two interior bulk
regions.

The thin-shell formalism of general
relativity~\cite{Israel,Lanczos,Sen} relates the discontinuity in
extrinsic curvature to the energy density and tension localized on the
junction:\footnote{
%----------------------------------------------------
The numerical coefficients appearing herein are dimension-dependent
(because of the implicit trace over the Ricci tensor and extrinsic
curvature hidden in the Einstein equations).}
%----------------------------------------------------
%
\begin{equation}
\sigma = -{1\over4\pi} \kappa_{\th\th} 
= \mp {1\over2\pi r}  \left.\sqrt{F(r)}\right|_{r=a}.
\end{equation}
\begin{equation}
\theta = -{1\over8\pi} \left[ \kappa_{\th\th} - \kappa_{\t\t} \right]
= \mp {1\over4\pi r}  
\left.{\partial\over\partial r} \left(r\sqrt{F(r)} \right)\right|_{r=a}.
\end{equation}
If the material located in the junction is a ``clean'' brane (a
brane in its ground state, without extra trapped matter in the form of
stringy excitations), then its equation of state is $\sigma=\theta$
and the condition for a static brane configuration (either a
wormhole or baby universe geometry) is simply
\begin{equation}
\sigma=\theta 
\qquad\implies\qquad  
2\left.\sqrt{F(r)}\right|_{r=a} = 
\left.{\partial\over\partial r} \left(r\sqrt{F(r)} \right)\right|_{r=a}
\qquad\implies\qquad
\left.{\partial\over\partial r} \left({F(r)\over r^2} \right)\right|_{r=a} = 0.
\end{equation}
Thus we have a very simple result: static wormholes (baby universes)
correspond to extrema of the function $F(r)/r^2$, though at this stage
we have not yet made any assertions about stability or dynamics. The
only difference between wormholes and baby universes is that for
wormholes the brane tension must be negative, whereas for baby
universes it is positive.

It is instructive to note that the locations of these static brane
solutions correspond to {\em circular photon orbits} in the original
spacetime (and this is true for arbitrary $F(r)$). That is: at these
static brane solutions any ``particle'' that is emitted form the
brane, which then follows null geodesics (of the bulk spacetime),
and which initially has no radial momentum, will just skim along the
brane; never moving off into the bulk. (Note that this is a purely
kinematic effect that occurs over and above any ``trapping'' due to
stringy interactions between the brane and excited string states.)

This may easily be verified by considering the photon orbits for
arbitrary $F(r)$. The time-translation and rotational Killing vectors
lead to conserved quantities
\begin{equation}
\left({\partial\over\partial t}\, , k\right) = -\epsilon
\qquad\implies\qquad  
g_{tt} \; {\d t\over \d\lambda} = -\epsilon
\qquad\implies\qquad  
F \; {\d t\over \d\lambda} = \epsilon.
\end{equation}
\begin{equation}
\left({\partial\over\partial \phi}\, , k\right) = -\ell
\qquad\implies\qquad  
g_{\phi\phi} \; {\d\phi\over \d\lambda} = \ell
\qquad\implies\qquad  
r^2 \; {\d\phi\over \d\lambda} = \ell.
\end{equation}
Inserting this back into the condition that the photon momentum be a
null vector, $(k,k)=0$, we see
\begin{equation}
\left({\d r\over \d\lambda}\right)^2 + {F(r) \; \ell^2\over r^2} = \epsilon^2.
\end{equation}
Now $\lambda$ is an arbitrary affine parameter, so we can
reparameterize $\lambda \to \ell \lambda$ and define $b=\epsilon/\ell$
to see that photon orbits are described by the equation
\begin{equation}
\left({\d r\over \d\lambda}\right)^2 + {F(r)\over r^2} = b^2.
\end{equation}
The circular photon orbits (and at this stage we make no claims about
stable versus unstable circular photon orbits) are, as claimed, at
extrema of the function $F(r)/r^2$ (which coincide with the location
of the stable brane configurations).

%----------------------------------------------------
\subsection{Brane dynamics}
%----------------------------------------------------
\label{SS:dynamics}
%----------------------------------------------------

If now the brane is allowed to move radially $a\to a(t)$, we start
the analysis by first parameterizing the motion in terms of proper
time along a curve of fixed $\theta$ and $\phi$. That is: the brane
sweeps out a world-volume
\begin{equation}
X^\mu(\tau,\theta,\phi) = \left(t(\tau),a(\tau),\theta,\phi\right).
\end{equation}
The 4-velocity of the $(\theta,\phi)$ element of the brane can then
be defined as
\begin{equation}
V^\mu = \left({\d t \over\d \tau}, {\d a\over\d\tau},0,0 \right).
\end{equation}
Using the normalization condition and the assumed form of the metric,
and defining $\dot a = \d a/\d\tau$,
\begin{equation}
V^\mu = 
\left({\sqrt{F(a)+\dot a^2}\over F(a)},\dot a, 0,0\right); 
\qquad 
V_\mu =
\left(-\sqrt{F(a)+\dot a^2},{\dot a\over F(a)},0,0\right).
\end{equation}
The unit normal vector to the sphere $a(\tau)$ is
\begin{equation}
n^\mu = 
\pm\left({\dot a\over F(a)},\sqrt{F(a)+ \dot a^2},0,0\right); 
\qquad 
n_\mu = 
\pm\left(-\dot a,{\sqrt{F(a)+\dot a^2}\over F(a)},0,0\right).
\end{equation}
The extrinsic curvature can still be written in terms of the normal
derivative
\begin{equation}
K_{\mu\nu} 
= \half \; n^\sigma \;  {\partial g_{\mu\nu}\over\partial x^\sigma}.
\end{equation}
If we go to an orthonormal basis, the $\th\th$ component is easily
evaluated~\cite{Book,Surgery}
\begin{equation}
K_{\th\th}
= \pm \half  \sqrt{F(a)+\dot a^2} \; {\partial g_{\theta\theta}\over\partial r} \; 
   g^{\theta\theta} 
= \pm {\sqrt{F(a)+\dot a^2}\over a}
\end{equation}
The $\tau\tau$ component is a little messier, but generalizing the
calculation of~\cite{Surgery} or~\cite{Book} (which amounts to
calculating the four-acceleration of the brane) quickly leads to\footnote{
%-------------------------------------------------------------------------
We do not repeat the details here since this calculation is now
standard textbook fare~\cite{Book}, pp 182--183. If one wishes to
avoid the need for this particular calculation one can instead work
backwards from the conservation of stress-energy, together with the
already-calculated expression for $K_{\th\th}$, to {\em deduce} an
expression for $K_{\hat{\tau}\hat{\tau}}$. But if you choose this
route you lose the opportunity to make a consistency check.}
%--------------------------------------------------------------------------
%
\begin{equation}
K_{\hat{\tau}\hat{\tau}} 
= \mp \half  {1\over\sqrt{F(a)+\dot a^2}} \; 
\left({\d F(r)\over\d a} + 2 \ddot a \right)
= \mp {\d\over\d a} \sqrt{F(a)+\dot a^2}.
\end{equation}
Applying the thin-shell formalism now gives:
\begin{equation}
\label{E1}
\sigma
= \mp {1\over2\pi a}  \; \sqrt{F(a)+\dot a^2}.
\end{equation}
\begin{equation}
\label{E2}
\theta
= \mp {1\over4\pi a}  
\; {\d\over\d a} \left(a \sqrt{F(r)+\dot a^2} \right).
\end{equation}
These equations can easily be seen to be compatible with the
conservation of the stress energy localized on the brane
\begin{equation}
\label{E3}
{\d\over\d\tau} (\sigma a^2) = \theta {\d\over\d\tau} (a^2).
\end{equation}
So as usual, {\em two} of these three equations are independent, 
and the third is redundant.

{From} the above we see that traversable wormhole solutions,
corresponding to the minus sign above, require negative brane tension
(and so positive internal pressure and negative internal energy
density). This is in complete agreement with~\cite{Hochberg} where it
was demonstrated that even for dynamical wormholes there must be
violations of the null energy condition at (or near) the throat.

If the material located in the junction is again assumed to be a
``clean'' brane ($\sigma=\theta$) then all the dynamics can be
reduced to a {\em single} equation~\footnote{
%--------------------------------------------
And if the brane is not ``clean'' in this sense one only needs to keep
track of one additional piece of information --- the on-brane
conservation equation (\ref{E3}).}
%--------------------------------------------
%
\begin{equation}
\dot a^2 + F(a) = (2\pi\sigma)^2 \; a^2.
\end{equation}
This single dynamical equation applies equally well to both wormholes
and baby universes, the $\mp$ that shows up in the brane Einstein
equations quietly goes away upon squaring --- thus for questions of
dynamics and stability these surgically constructed baby universes and
wormholes can be dealt with simultaneously --- the {\em only} difference
lies in question of whether the brane tension is positive or negative.

Note that we could re-write this dynamical equation as
\begin{equation}
\left({\d \ln(a) \over\d\tau}\right)^2 + {F(a)\over a^2} = (2\pi\sigma)^2.
\end{equation}
{From} this it is clear that static solutions must be located at extrema
of the function $F(a)/a^2$, in agreement with the static analysis.

In the next section we shall make use of this general formalism by
specializing $F(r)$ to the {\RNdS} form.  We shall then exhibit some
explicit solutions to these brane equations of motion, and perform
the relevant stability analysis.

%------------------------------------------------------------------------------
\section{Reissner--\Nordstrom--de~Sitter surgery}
%------------------------------------------------------------------------------
\label{S:RNdS}
%------------------------------------------------------------------------------

For the {\RNdS} geometry
\begin{equation}
F(r) = 1 - {2M\over r} + {Q^2\over r^2} - {\Lambda\over 3}\; r^2.
\end{equation}
It is then most instructive to write the dynamical equation in the form
\begin{equation}
\left({\d \ln(a) \over\d\tau}\right)^2 + V(a) = E,
\end{equation}
with a ``potential''
\begin{equation}
V(a) = {F(a)\over a^2} 
=  {1\over a^2} - {2M\over a^3} + {Q^2\over a^4} - {\Lambda\over3},
\end{equation}
and an ``energy''
\begin{equation}
E =  + (2\pi\sigma)^2.
\end{equation}
The extrema of this potential are easily located, their positions are
independent of $\Lambda$ and occur at
\begin{equation}
r_\pm = {3M\over2} \pm \sqrt{\left({3M\over2}\right)^2 - 2 Q^2 }.
\end{equation}
(As promised, these are indeed the locations of the circular photon
orbits of the {\RNdS} geometry; note that the cosmological constant
does {\em not\/} affect the location of these circular photon orbits.)
The {\em value} of this potential at these extrema is somewhat tedious
to calculate, we find
\begin{eqnarray}
V_\pm(M,Q,\Lambda) &\equiv& V(r_\pm(M,Q)) 
\\
&=& 
-{1\over4Q^2}\left(1-{9\over2}{M^2\over Q^2}+{27\over8}{M^4\over Q^4}\right)
\pm {M\over4Q^6} \left[\left({3M\over2}\right)^2-2Q^2\right]^{3/2}
- {\Lambda\over3}.
\nonumber
\end{eqnarray}
Though it may not be obvious, the $Q\to 0$ limit formally exists and
is given by
\begin{equation}
V_-(M,Q\to0,\Lambda) \to -\infty; \qquad 
V_+(M,Q\to0,\Lambda) \to {1\over 27 M^2} - {\Lambda\over3}.
\end{equation}

\noindent
The behaviour of the potential $V(a)$ is qualitatively:
\begin{itemize}
\item
$V(a)\to+\infty$ as $a\to 0$, ($Q\neq0$).
\item
$V(a)\to - {\Lambda\over3}$ as $a\to +\infty$.
\item
There is at most one local minimum ($V_-$ located at $r_-$) and one
local maximum ($V_+$ located at $r_+$).
\end{itemize}
Two figures, where we have plotted $V(a)$ for two special cases, are
provided in the discussion below.  When looking for a stable brane
solution we want to satisfy the following:
\begin{enumerate}
\item
We want the local minimum to exist, and the brane to be located in its
basin of attraction.
\item
The energy must be at least equal to $V_-$ (to even get a solution),
and should not exceed $V_+$ (to avoid having the solution escape from
the local well located around $r_-$).
\item
We also do not want (at least for now) the brane to fall inside (or
even touch) any horizon the original {\RNdS} geometry might have ---
for two reasons:
\begin{enumerate}
\item
Because if it does fall inside (or even touch) an event horizon the
wormhole geometry is operationally indistinguishable from a {\RNdS}
black hole and therefore not particularly interesting (but see the
discussion regarding singularity avoidance later in this paper)
whereas the baby universe geometry is for $Q=0$ doomed to a brief and
unhappy life, and for $Q\neq0$ is just plain weird.
\item
For technical reasons ($r$ is now timelike and $t$ spacelike) a few
key minus signs flip at intermediate steps of the calculation, more on
this later.
\end{enumerate}
\end{enumerate}

\noindent
These physical constraints now imply:
\begin{enumerate}
\item
To get a local minimum we need $M > \sqrt{8/9} \; Q$.
\item
To then trap the solution, to make it one of bounded excursion, we need
\begin{equation}
V_-(M,Q,\Lambda) \leq + (2\pi\sigma)^2 \leq V_+(M,Q,\Lambda).
\end{equation}
\item
Horizon avoidance requires $F(a)>0$ over the entire range of motion;
this implies 
\begin{equation}
V(a) = {F(a)\over a^2} > 0;
\qquad\implies\qquad
V_-(M,Q,\Lambda) > 0.
\end{equation}
In view of this the horizon avoidance condition might more properly be
called horizon elimination --- horizons can be avoided if and only if
the inner and outer horizons are actually eliminated. (We could
however still have a cosmological horizon at very large distances,
this cosmological horizon is never reached if the bounded excursion
constraint is satisfied.) We can also explicitly separate out the
cosmological constant to write the horizon elimination condition as
\begin{equation} 
\Lambda < 3 \; V_-(M,Q,\Lambda\to0),
\end{equation}
which makes it clear that a powerful enough negative (bulk)
cosmological constant is guaranteed to eliminate all the event
horizons from the geometry.
\end{enumerate}
That these constraints can simultaneously be satisfied (at least in
certain parameter regimes) can now be verified by inspection. The best
way to proceed is to sub-divide the discussion into several special
cases.

%------------------------------------------------------------------------------
\subsection{$M>|Q|=0$:}
%------------------------------------------------------------------------------

There is one maxima (at $a=3M$) and no minimum. There are no stable
solutions, though the ``arbitrarily advanced civilization'' posited by
Morris and Thorne~\cite{Morris-Thorne} might like to try to artificially
maintain the unstable static solution at $a=3M$. (This solution is
unstable to both collapse and explosion.)

Adding $Q\neq0$ provides a hard core to the potential so that collapse
is avoided (modulo the horizon crossing issue which must be dealt with
separately).

%------------------------------------------------------------------------------
\subsection{$M>|Q|\neq0$:}
%------------------------------------------------------------------------------

\begin{figure}[htb]
\vbox{ 
\hfil
\scalebox{0.75}{\includegraphics{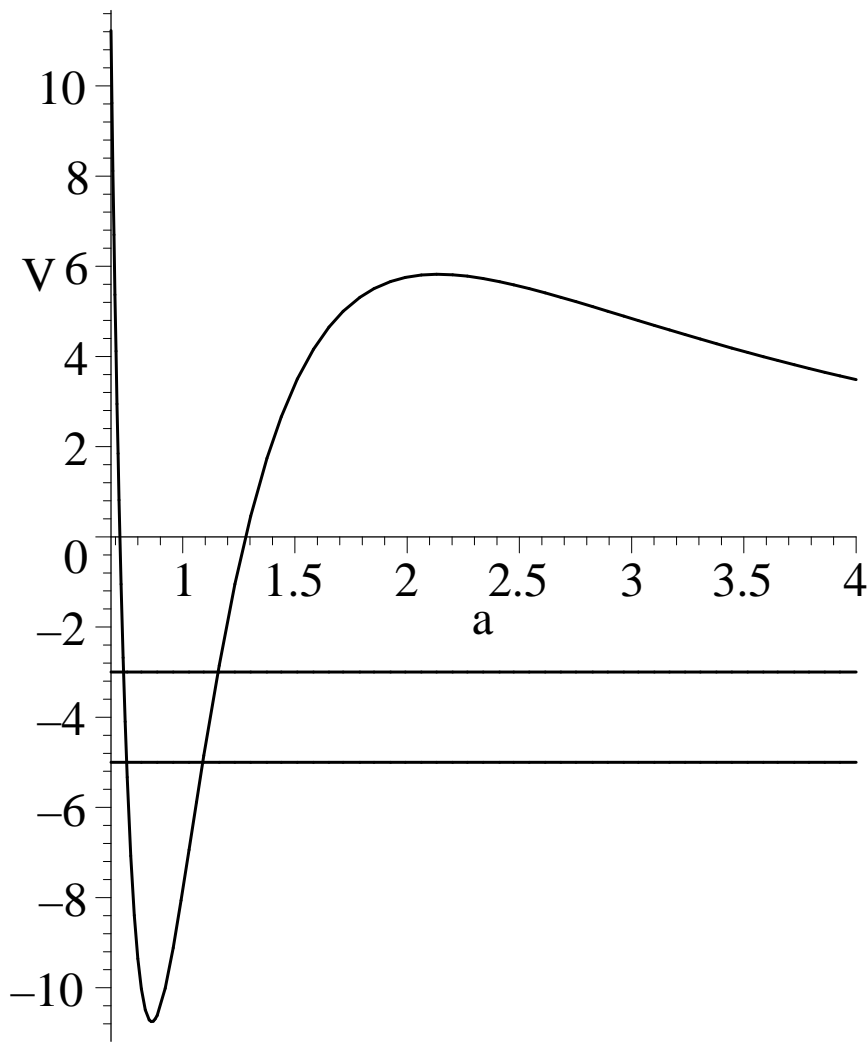}}
\hfil 
}
\bigskip
\caption{\label{F:1} 
Sketch of the potential $V(a)$ for $M>|Q|$ and $\Lambda=0$.  Adding a
cosmological constant merely moves the entire curve up or down: the
lower horizontal line represents $\Lambda/3$, and for $\Lambda$
sufficiently large and negative the inner and outer horizons (which
are given by the intersection of this horizontal line with the
$\Lambda=0$ potential) are guaranteed to be eliminated. The upper
horizontal line represents $\Lambda/3 + (2\pi\sigma)^2$, and its
intersection with the $\Lambda=0$ potential gives the turning points
of the motion. If inner and outer horizons exist they lie between the
inner and outer turning points.}
\end{figure}

There are now both a local maximum (at $r_+<3M$) and a global minimum
(at $r_->0$).  The potential is plotted in figure \ref{F:1}. Stable
solutions exist, (both static stable solutions and stable solutions of
bounded excursion), but since $V_-<0$ ($\Lambda=0$) at the global
minimum horizon avoidance requires
\begin{equation}
\Lambda < 3 \; V_-(M,Q,\Lambda\to0) < 0.
\end{equation}
That is, stable traversable wormhole or baby universe solutions exist
only if the bulk is anti-de~Sitter space (adS${}_{(3+1)}$) with a strong
enough negative cosmological constant.  

Indeed if you consider the original geometry prior to brane surgery
and extend it down to $r=0$ then for this choice of parameters
(because of the large negative cosmological constant) you encounter a
naked singularity. For the stable wormhole geometries based on this
brane prescription this is {\em not\/} a problem since the naked
singularity was in the part of the spacetime that you threw away in
setting up the brane construction. (The baby universe models on the
other hand, while stable, explicitly do contain naked
singularities.)\footnote{
%------------------------------------------------------------------
This is part of a general pattern: The stable (or even merely static)
brane configurations that do not possess naked singularities in the
bulk region are the wormhole configurations with negative brane
tension. This observation also applies to the other sub-cases
discussed below. This is compatible with the discussion of Chamblin,
Perry, and Reall~\cite{Reall} who discovered qualitatively similar
behaviour for (8+1)-dimensional branes in a (9+1)-dimensional bulk.
Specifically, they found that static (8+1)-dimensional brane
configurations with positive brane tension led to naked singularities
in the bulk, and that eliminating the naked singularities forced the
adoption of negative brane tension (and implicitly a wormhole
configuration). This observation also serves to buttress our previous
comments to the effect that the qualitative features of the
calculations presented in this paper are generic, and are not just
limited to (2+1) branes in (3+1) dimensions.}
%----------------------------------------------------------------------

A particularly simple sub-class of these solutions occurs when the
bulk cosmological constant is tuned to a special value in terms of the
brane tension. This is the analog of the Randall--Sundrum fine
tuning~\cite{RS1,RS2} and corresponds to a zero ``effective''
cosmological constant, in the sense that the brane equation of motion
can be rearranged and reinterpreted as being governed by $E=0$ and
\begin{equation}
\Lambda_\effective = \Lambda + 3(2\pi\sigma)^2.
\end{equation}
If this $\Lambda_\effective$ is now tuned to zero
\begin{equation}
\Lambda = -3 (2\pi\sigma)^2 < 3 \; V_-(M,Q,\Lambda\to0) < 0.
\end{equation}

%==================================================== 
%====================================================

%------------------------------------------------------------------------------
\subsection{$M=|Q|$:}
%------------------------------------------------------------------------------

There are still both a local maximum (at $r_+=2M$) and a global
minimum (at $r_-=M$).  Stable solutions exist. Since now
$V_-(\Lambda\to0) = 0$at the global minimum horizon avoidance requires
anti de~Sitter space with an arbitrarily weak cosmological
constant. (And again this is an example of horizon elimination.)

%------------------------------------------------------------------------------
\subsection{$M\in(\sqrt{8/9}\;|Q|,|Q|)$:}
%------------------------------------------------------------------------------

\begin{figure}[htb]
\vbox{ 
\hfil
\scalebox{0.75}{\includegraphics{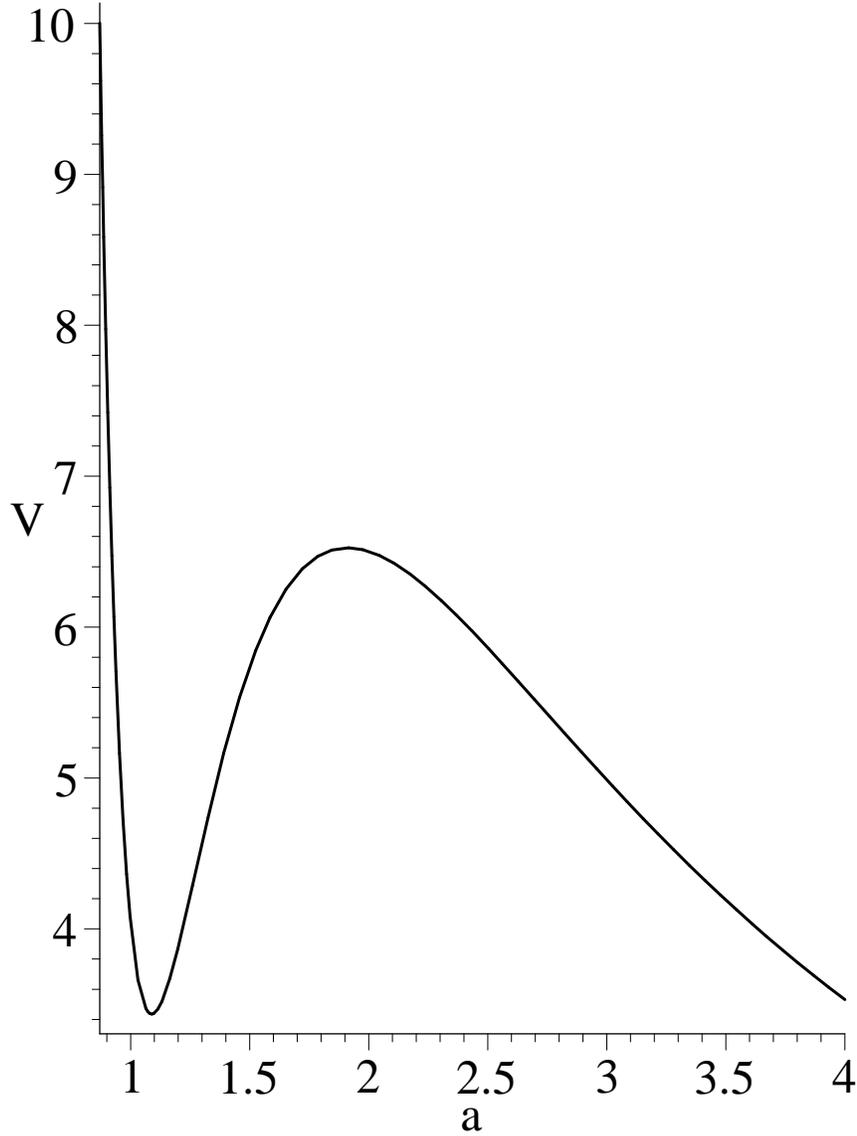}}
\hfil 
}
\bigskip
\caption{\label{F:2} 
Sketch of the potential $V(a)$ for $M$ in the critical range
$(\sqrt{8/9}\;|Q|, \; |Q|)$, and $\Lambda=0$.  Adding a cosmological
constant merely moves the entire curve up or down. In this case, even
for $\Lambda=0$, we see a stable minimum at $r_-$ with no event
horizons. For small positive $\Lambda$ a cosmological horizon will
form at very large radius, but this is of no immediate concern because
of the barrier at $r_+$. If $\Lambda$ becomes too large however,
$\Lambda > V_-(M,Q,\Lambda\to0)$, inner and outer horizons will
reappear between the inner and outer turning points.}
\end{figure}

There are still both a local maximum (at $r_+<2M$) and a global
minimum (at $r_->M$).  The potential is plotted in figure \ref{F:2}.
Stable solutions exist. Since now $V_-(\Lambda\to0) > 0$ at the global
minimum, horizon avoidance can be achieved with zero cosmological
constant in the bulk. For instance, picking
\begin{equation}
\Lambda = 0; 
\qquad
(2\pi\sigma)^2 = V_-(M,Q,\Lambda\to0),
\end{equation}
yields the stable static solution at $r_-$. This is perhaps the most
``physical'' of these traversable wormholes in that it resides in an
asymptotically flat spacetime. 

%==================================================== 
%====================================================

%------------------------------------------------------------------------------
\subsection{$M=\sqrt{8/9}\; |Q|$:}
%------------------------------------------------------------------------------

The maximum and minimum merge into a single point of inflection (at
$r_\pm = 3M/2$). There are no stable solutions. All the solutions
exhibit runaway to large radius.

%------------------------------------------------------------------------------
\subsection{$M\in(\sqrt{8/9}|Q|,0)$:}
%------------------------------------------------------------------------------

There is not even a point of inflection: the potential is monotonic
decreasing. There are no stable solutions. 

%------------------------------------------------------------------------------
\subsection{$M=0$; $Q\neq0$:}
%------------------------------------------------------------------------------

There is not even a point of inflection: the potential is monotonic
decreasing. There are no stable solutions. 

%------------------------------------------------------------------------------
\subsection{$M<0$:}
%------------------------------------------------------------------------------

Letting the central mass $M$ go negative is not helpful --- $M<0$
helps stabilize against collapse, but actually destroys the
possibility of stable solutions because the location of ``extrema''
$r_\pm$ is pushed to unphysical nominally negative values of the
radius.

%------------------------------------------------------------------------------
\subsection{Baby bangs?}
%------------------------------------------------------------------------------

The fact that so many of these baby universe models are unstable to
explosion is intriguing, and potentially of phenomenological
interest. While these particular baby-universe models are not suitable
cosmologies for our own universe, we believe that more realistic
scenarios can be developed.

%------------------------------------------------------------------------------
\subsection{Singularity avoidance?}
%------------------------------------------------------------------------------

We have so far sought to implement horizon avoidance in our models: we
have sought conditions that would prevent the brane from falling
through or even touching any horizon that might be present in the
underlying pre-surgery spacetime. Suppose we now relax that
constraint. The best way to analyze the situation is to note that
inside the horizon (more precisely between the outer horizon and the
inner horizon) the pre-surgery metric can be written in the form
\begin{equation}
\d s^2 = + |F(r)| \; \d t^2 -{\d r^2\over |F(r)|} + r^2 \; \d\Omega_2^2.
\end{equation}
The calculation of the four-velocity, normal, extrinsic curvatures,
and their discontinuities can be repeated, with the result that in this
region [$F(r)<0$]
\begin{equation}
V^\mu = \left(-{\sqrt{\dot a^2-|F(a)|}\over |F(a)|},\dot a, 0,0\right); 
\qquad 
n^\mu = \pm\left(-{\dot a\over |F(a)|},\sqrt{\dot a^2-|F(a)|},0,0\right); 
\end{equation}
and
\begin{equation}
\sigma = \mp {1\over2\pi a}  \; \sqrt{\dot a^2 - |F(a)|}.
\end{equation}
After rearrangement this leads to the {\em same} dynamical equation as
before
\begin{equation}
\left({\d \ln(a) \over\d\tau}\right)^2 + {F(a)\over a^2} = (2\pi\sigma)^2.
\end{equation}
So that all of our previous arguments can be extended inside the event
horizon.

\begin{figure}[htb]
\vbox{ 
\hfil
\scalebox{\SIZE}{\includegraphics{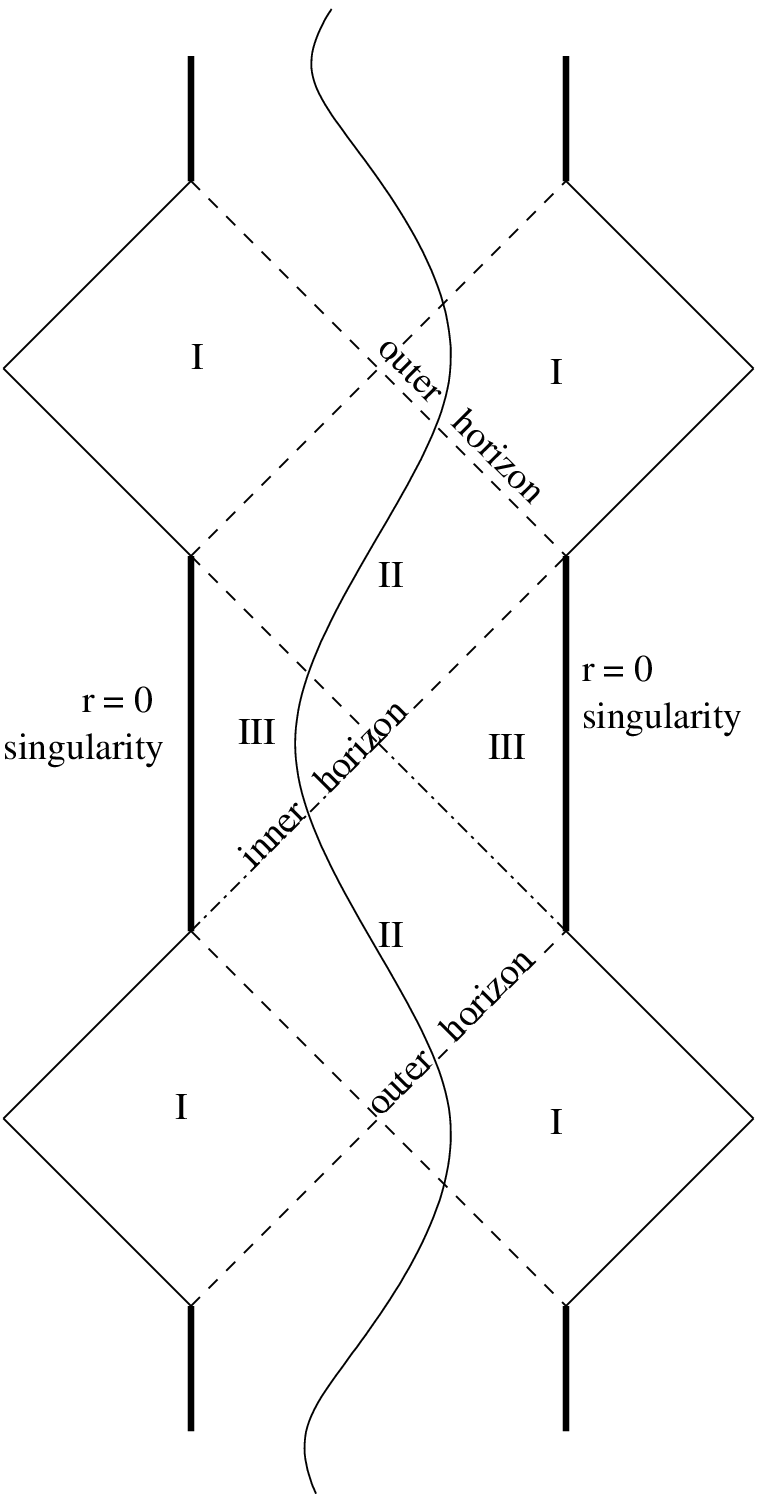}}
\hfil 
}
\bigskip
\caption{\label{F:3} 
Sketch of the Penrose diagram for the maximally extended
Reissner-{\Nordstrom} geometry when $M>|Q|$ ($\Lambda = 0$). A
timelike (2+1)-brane [spacelike normal] will oscillate between the
turning points $r_+$ and $r_-$, but each oscillation will take
infinite coordinate time even if it takes finite proper time. For
wormhole solutions keep the right half of the diagram, make two
copies, and sew them together along the brane. For baby universe
geometries keep the left half of the diagram, make two copies and sew
them up along the brane.}
\end{figure}

\noindent
A few key observations:
\begin{itemize}
\item
The two turning points occur at $F(a)/a^2 = (2\pi\sigma)^2 > 0$. Thus
$F(a) > 0$ at the turning points. So if there are horizons present
(that is, if $F(a)=0$ has solutions $r_{\mathrm{horizon}}^\pm \neq
r_\pm$), and one is in the potential well near $r_-$, then one turning
point will be outside the outer horizon, and the second turning point
will be inside the inner horizon.
\item
Even though the brane oscillation will take finite proper time $\tau$
this corresponds to infinite $t$-parameter time --- when the brane
re-emerges from the outer horizon it will emerge from a past outer
horizon of a ``future incarnation'' of the universe; the brane will
not re-emerge into our own universe. (For simplicity you may wish to
set $\Lambda=0$ and consider the Penrose diagram of the maximally
extended Reissner--{\Nordstrom} geometry as presented, for instance,
on page 158 of Hawking and Ellis~\cite{Hawking-Ellis}. A partial
Penrose diagram for {\RNdS} may be found in~\cite{Poisson}. See also
figure \ref{F:3}.)
\item
Operationally, from ``our'' asymptotically flat region, once the brane
passes the horizon the geometry will be indistinguishable from an
ordinary {\RNdS} black hole.
\item
The original pre-surgery spacetime has two asymptotic regions, two
outer horizons, and two inner horizons, which are then repeated an
infinite number of times in the maximal analytic extension. If the
brane starts out in the rightmost asymptotic region and falls through
the right (future) outer horizon, then you can quickly convince
yourself that it must pass through the {\em left} inner horizon
(twice, once on the way in, and once again on the rebound) before
moving back out through the right (past) outer horizon back into the
(next incarnation of) the right asymptotic region.  (See figure
\ref{F:3}.)
\item
The wormhole geometry based on this brane surgery is an explicit
example of {\em partial} evasion of the usual singularity
theorems~\cite{Hawking-Ellis}. (We say evasion, not violation, because
the presence of the negative tension brane vitiates the usual
hypotheses used in proving the singularity theorems.) The wormhole
geometry certainly has trapped surfaces once the brane falls inside
the horizon, but by construction there is no {\em left} curvature
singularity. (The {\em right} curvature singularity is still there,
and the right inner horizon is still a Cauchy horizon.)\footnote{
%---------------------------------------------------
If you think of the {\RNdS} geometry as arising from gravitational
collapse of an electrically charged star, then it is the left curvature
singularity (which is eliminated by the present construction) that
would arise from the central density of the star growing to
infinity. The right curvature singularity (which is unaffected by the
present construction) has a totally different genesis as it arises in
a matter-free region due to gravitational focussing of the
electromagnetic field.}
%---------------------------------------------------
Note that this is a idealized statement appropriate to ``clean''
wormhole universes containing only a few test particles of matter: in
any more realistic model where the universe contains a finite amount
of radiation, inner event horizons are typically unstable to a violent
blue shift instability, and are typically converted by back-reaction
effects to some sort of curvature singularity~\cite{Poisson}. This
process however, lies far beyond the scope of the usual singularity
theorems.
\end{itemize}
%

%==================================================== 
%====================================================

If you wish to eliminate {\em both} left and right singularities a
more drastic fix is called for: You will need to use a
(3+0)-dimensional brane, something you might call an instanton-brane
because it represents a spacelike hypersurface through the spacetime
--- at early times there's nothing there, the brane ``switches on''
for an instant, and then it's gone again. The simplest example of such
a instanton-brane is to place one at $r_-$, the static minimum of the
potential $V(a)$.\footnote{
%-----------------------------------------------------------------
Although this is a static minimum of the usual $V(a)$ it is in the
present context not {\em stable}. This arises because for a spacelike
shell the overall {\em sign} of the potential flips.}
%------------------------------------------------------------------
If there are event horizons then this minima will be inside the event
horizon (between inner and outer horizons) and a hypersurface placed
at $r_-$ will be spacelike. Placing the instanton-brane at this
location will eliminate {\em both} singularities and {\em both} inner
horizons --- you are left with two asymptotic regions and two (outer)
event horizons, infinitely repeated.

More generally one could think of an instanton-brane described by a
location $a(\ell)$ where $\ell$ is now proper length along the brane
(and the notion of {\em dynamics} is somewhat obscure). The spacelike
tangent and timelike normal are now (outside the horizon)
\begin{equation}
V^\mu = \left({\sqrt{(\d a/\d\ell)^2-F(a)}\over F(a)},
{\d a\over\d\ell}, 0,0\right); 
\qquad 
n^\mu = \pm\left({1\over F(a)} {\d a\over\d\ell},
\sqrt{(\d a/\d\ell)^2-F(a)},0,0\right); 
\end{equation}
and a brief computation yields
\begin{equation}
\sigma = \mp {1\over2\pi a}  \; \sqrt{(\d a/\d\ell)^2-F(a)}.
\end{equation}
This can be rearranged to give
\begin{equation}
\left({\d \ln(a) \over\d\ell}\right)^2 - {F(a)\over a^2} = (2\pi\sigma)^2.
\end{equation}
So the net result is that for an instanton-brane the {\em sign} of the
potential has flipped, but that of the brane contribution to the
energy has not. (And exactly the same result continues to hold inside
the horizon, a few intermediate signs flip, but that's all.)

In summary: certain varieties of brane wormhole provide explicit
evasions (either partial or complete) of the usual singularity
theorems.

%------------------------------------------------------------------------------
\section{Voids: the brane as a spacetime boundary}
%------------------------------------------------------------------------------
\label{S:voids}
%------------------------------------------------------------------------------

A somewhat unusual feature of brane physics is that the brane
could also be viewed as an actual physical boundary to spacetime, with
the ``other side'' of the brane being null and void. In general
relativity as it is normally formulated the notion of an actual
physical boundary to spacetime (that is, an accessible boundary
reachable at finite distance) is anathema. The reason that spacetime
boundaries are so thoroughly deprecated in general relativity is that
they become highly artificial special places in the manifold where
some sort of boundary condition has to be placed on the physics by an
act of black magic. Without such a postulated boundary condition all
predictability is lost, and the theory is not physically
acceptable. Since there is no physically justifiable reason for
picking any one particular type of boundary condition (Dirichlet,
Neumann, Robin, or something more complicated), the attitude in
standard general relativity has been to exclude boundaries, by
appealing to the cosmic censor whenever possible and by hand if
necessary.

The key difference when a brane is used as a boundary is that now
there is a specific and well-defined boundary condition for the
physics: D-branes ({\em D for Dirichlet\/}) are defined as the loci on
which the fundamental open strings end (and satisfy Dirichlet-type
boundary conditions). D-branes are therefore capable (at least in {\em
principle}) of providing both a physical boundary {\em and} a
plausible boundary condition for spacetime. For Neveu--Schwarz branes
the boundary conditions imposed on the fundamental string states are
more complicated, but they still (at least in principle) provide
physical boundary conditions on the spacetime.

When it comes to specific calculations, this may however not be the
best mental picture to have in mind --- after all, how would you try
to calculate the Riemann tensor for the edge of spacetime? And what
would happen to the Einstein equations at the edge? There is a
specific trick that clarifies the situation: Take the manifold with
brane boundary and make a second copy, then sew the two manifolds
together along their respective brane boundaries, creating a single
manifold without boundary that contains a brane, and exhibits a
$Z_2$ symmetry on reflection around the brane. Because this new
manifold is a perfectly reasonable no-boundary manifold containing a
brane, the gravitational field can be analyzed using the usual
thin-shell formalism of general relativity~\cite{Israel,Lanczos,Sen}:
The metric is continuous, the connection exhibits a step-function
discontinuity, and the Riemann curvature a delta-function at the
brane. The dynamics of the brane can then be investigated in this
$Z_2$-doubled manifold, and once the dynamical equations and their
solutions have been investigated the second surplus copy of spacetime
can quietly be forgotten.

In particular, all the calculations we have performed for the
spherically symmetric wormholes of this paper apply equally well to
spherically symmetric holes in spacetime (not black holes, actual
voids in the manifold), with the edge of the hole being a brane ---
we deduce the existence of a large class of stable void solutions, and
an equally large class of unstable voids that either collapse to form
black holes, or explode to engulf the entire universe.

Equally well, the baby universes of the preceding section can, under
this new physical interpretation of the relevant mathematics, be used
to investigate finite volume universes with boundary. The bulk of the
physical universe now lies in the range $r\in(0,a)$, and the edge of
the universe is located at $a$. Again, we deduce the existence of a
large class of stable baby universes with boundary, and an equally
large class of unstable baby universes that either collapse to
singularity, or explode to provide arbitrarily large universes. Note
that these particular exploding universes are {\em not\/} FLRW
universes, and are not suitable cosmologies for our own universe.
Nevertheless, this notion of using a brane as an actual physical
boundary of spacetime is an issue of general applicability, and we
hope to return to this topic in future publications.

%------------------------------------------------------------------------------
\section{Discussion}
%------------------------------------------------------------------------------
\label{S:discusion}
%------------------------------------------------------------------------------

The main point of this paper is that in the brane picture there is
nothing wrong with the notion of a {\em negative} brane tension, and
that once branes of this type are allowed to contribute to the
stress-energy, the class of solutions is greatly enhanced, now
including many quite peculiar beasts not normally considered to be
part of standard general relativity.  As specific examples, the energy
condition violations caused by negative tension branes allow one to
construct classical traversable wormholes, at least some of which (as
we have seen) are actually dynamically stable. Now for spherically
symmetric wormholes of the type considered in this paper, attempting
to cross from one universe to the other requires the traveller to
cross the brane, a process which is likely to prove disruptive of
the traveller's internal structure, well being, and overall
health. This problem, or rather the no-brane analog of this problem,
was already considered by Morris and Thorne in their pioneering work
on traversable wormholes~\cite{Morris-Thorne}. A possible resolution
comes from the fact that spherical symmetry is a considerable
idealization: One of the present authors has demonstrated that if one
uses negative tension cosmic strings instead of negative tension
domain walls, then it is possible to construct traversable wormhole
spacetimes that do not possess spherical symmetry, and contain
perfectly reasonable paths from one asymptotic region to the other
that do not involve personal encounters with any form of ``exotic
matter''~\cite{Examples}. (See also the extensive discussion
in~\cite{Book}.) In a brane context this means we should consider the
possibility of a negative tension (1+1)-dimensional brane in
(3+1)-dimensional spacetime.

Now the peculiarities attendant on widespread violations of the energy
conditions are not limited to violations of topological censorship; as
we have seen there is also the possibility of violating (evading) the
singularity theorem. If this is not enough, then it should be borne in
mind that without some form of energy condition we do not have a
positive mass {\em theorem}. (Looking out into our own universe, we do
have a positive mass {\em observation}, but it would be nice to be
able to deduce this from general principles.) A discussion of some of
the peculiarities attendant on negative asymptotic mass can be found
in the early work of Bondi~\cite{Bondi}, and a possible observational
signal (particular types of caustics in the light curves due to
gravitational lensing) has been pointed out by Cramer
{\etal}~\cite{Cramer}. Finally, energy condition violations are also
the {\em sine qua non} for the Alcubierre ``warp
drive''~\cite{Alcubierre}.

In summary, all of these somewhat peculiar geometries, which were
investigated within the general relativity community more with a view
to understanding the limitations of general relativity (and more
specifically, of semiclassical general relativity) than in the
expectation that they actually exist in reality, are now seen to
automatically be part and parcel of the brane models currently being
considered as semi-phenomenological models of empirical reality.

%------------------------------------------------------------------------------
\section*{Acknowledgments}
%------------------------------------------------------------------------------

The research of CB was supported by the Spanish Ministry of Education
and Culture (MEC). MV was supported by the US Department of Energy. We
wish to thank Harvey Reall and Sumit Das for their comments and
interest.

%------------------------------------------------------------------------------
 
%------------------------------------------------------------------------------
\end{document}